\documentclass[10pt,journal,compsoc]{IEEEtran}
\usepackage{latexsym}
\usepackage{amsfonts}
\usepackage{amsmath}
\usepackage{amssymb}
\usepackage{color}
\usepackage{epsfig}
\usepackage{xspace}
\usepackage{graphicx}
\usepackage{subfigure}
\usepackage{cite}
\usepackage[english]{babel}
\usepackage{balance}  

\usepackage{epsfig}
\usepackage{multirow}
\usepackage{url}

\usepackage{enumitem}
\setlist{topsep=0pt,noitemsep} \setitemize[1]{label=$\circ$}

\newcommand{\eat}[1]{}

\newcommand{\bi}{\begin{itemize}}
\newcommand{\ei}{\end{itemize}}

\newcommand{\be}{\begin{enumerate}}
\newcommand{\ee}{\end{enumerate}}
\newcommand{\beqn}{\begin{eqnarray*}}
\newcommand{\eeqn}{\end{eqnarray*}}

\newcommand{\stitle}[1]{\vspace{1ex}\noindent{\bf #1}}

\newcommand{\ie}{\emph{i.e.,}\xspace}
\newcommand{\eg}{\emph{e.g.,}\xspace}
\newcommand{\wrt}{\emph{w.r.t.}\xspace}
\newcommand{\aka}{\emph{a.k.a.}\xspace}

\newcommand{\kw}[1]{{\ensuremath {\mathsf{#1}}}\xspace}

\newcounter{ccc}

\newcommand{\CFDconsistency}{{\mbox{\s
\documentclass{sig}
\usepackage{algorithm}
\usepackage[noend]{algorithmic}
\usepackage{latexsym}
\usepackage{amsfonts}
\usepackage{amsmath}
\usepackage{amssymb}
\usepackage{color}
\usepackage{epsfig}
\usepackage{xspace}
\usepackage{graphicx}
\usepackage{times}
\usepackage{subfigure}mall\sf CFD\_Checking}\xspace}}

\newcommand{\NP}{{\sc np}\xspace}

\newcommand{\eop}{\hspace*{\fill}\mbox{$\Box$}}     
\newcounter{example}
\renewcommand{\theexample}{\arabic{example}}

\renewcommand{\ni}{\noindent}

\newcommand{\nthesection}{\arabic{section}}
\newcounter{theorem}
\renewcommand{\thetheorem}{\arabic{theorem}}
\newcounter{prop}
\renewcommand{\theprop}{\arabic{theorem}}
\newcounter{lemma}
\renewcommand{\thelemma}{\arabic{theorem}}
\newcounter{cor}
\renewcommand{\thecor}{\arabic{theorem}}
\newcounter{definition}[section]
\renewcommand{\thedefinition}{\nthesection.\arabic{definition}}

\newcounter{alg}[section]
\renewcommand{\thealg}{\nthesection.\arabic{alg}}

\newcounter{arule}
\renewcommand{\thearule}{\arabic{arule}}

\newcounter{claim}
\renewcommand{\theclaim}{\arabic{claim}}

\renewcommand{\texttt}[1]{{\small\textsf{#1}}}

\newcommand{\dist}{\kw{dist}}

\definecolor{gray}{rgb}{0.5,0.5,0.5}

\newcommand{\lsa}{\kw{LS}}

\newcommand{\operb}{\kw{OPERB}}
\newcommand{\operba}{\kw{OPERB}-\kw{A}}

\newcommand{\fbqsa}{\kw{FBQS}}
\newcommand{\dpa}{\kw{DP}}

\renewcommand{\path}[1]{{\sc path}${\kw{#1}}$}
\renewcommand{\dist}[1]{{\sc dist}${\kw{#1}}$}
\newcommand{\dra}{{\sc dra}\xspace}
\newcommand{\dras}{{\sc dra}s\xspace}

\newcommand{\ah}{{\sc ah}\xspace}

\title{Approximate Computation Toward Big Data Analytics}
\title{Approximate Computation for Big Data Analytics}

\author{Shuai~Ma, Jinpeng~Huai

\IEEEcompsocitemizethanks{\IEEEcompsocthanksitem \vspace{-1ex} S.~Ma and J.~Huai are with the SKLSDE Lab \& Beijing Advanced Innovation Center for Big Data and Brain Computing, Beihang University, Beijing, China.\protect\\
E-mail: \{mashuai,  huaijp\}@buaa.edu.cn. \vspace{-1.5ex}}
}

\date{}
\begin{document}

\IEEEtitleabstractindextext{
\begin{abstract}
Over the past a few years, research and development has made significant progresses on big data analytics.
A fundamental issue for big data analytics is the efficiency. If the optimal solution is unable to attain or not required or has a price to high to pay, it is reasonable to sacrifice optimality with a ``good'' feasible solution that can be computed efficiently. Existing approximation techniques can be in general classified into approximation algorithms,  approximate query processing for aggregate SQL queries and approximation computing for multiple layers of the system stack.
In this article, we systematically introduce approximate computation, \ie query approximation and data approximation, for efficiency and effectiveness big data analytics. We first explain the idea and rationale of query approximation,  and show efficiency can be obtained with high effectiveness in practice with three analytic tasks: graph pattern matching, trajectory compression and dense subgraph computation. We then explain the idea and rationale of data approximation,  and show efficiency can be obtained even without sacrificing for effectiveness in practice  with three analytic tasks: shortest paths/distances, network anomaly detection and link prediction.
\end{abstract}

}

\maketitle

\section{Introduction}
\label{sec-intro}

Over the past a few years, research and development has made significant progresses on big data analytics with the supports from both governments and industries all over the world, such as Spark\footnote{\small \url{https://spark.apache.org}}, IBM Watson\footnote{\small \url{https://www.ibm.com/watson}} and Google AlphaGo\footnote{\small \url{https://deepmind.com/research/alphago}}. A fundamental issue for big data analytics is the efficiency, and various advances towards attacking this issue have been achieved recently, from theory to algorithms to systems~\cite{FanGN13,Jordan15,ZahariaXWDADMRV16}. However, {\em if the optimal solution is unable to attain or not required or has a price to high to pay, it is reasonable to sacrifice optimality with a ``good'' feasible solution that can be computed efficiently}. Hence, various approximation techniques have been developed, and can in general be classified into three aspects: algorithms, SQL aggregate queries and multiple layers of the system stack.

\bi
\item[(1)] {\em Approximation algorithms} were formally defined in the 1970s~\cite{GareyGU72,Johnson74a}. An approximation algorithm is necessarily polynomial, and is evaluated by the worst case possible relative error over all possible instances of the  NP-hard optimization problem, under the widely believed $P\ne NP$ conjecture. This is relatively mature research field algorithm community, many approximation algorithm have been designed for optimization problems (see books~\cite{Dorit96,approx03,Ausiello99}).
\item[(2)] {\em Approximate query processing} supports a slightly constrained set of SQL-style declarative queries, and it specifically provides approximate results for standard SQL aggregate queries, \eg queries involving COUNT, AVG, SUM and PERCENTILE. Over the past two decades, approximate query processing has been successfully studied, among which sampling technique are heavily employed~\cite{ChaudhuriDK17,Mozafari17,Kraska17,GarofalakisG01}. Not only traditional DBMS systems, such as Oracle\footnote{\small \url{https://oracle-base.com/articles/12c/approximate-query} \url{-processing-12cr2}},  provide approximate functions to support approximate results, but also emerging new systems specially designed for approximate queries, such as BlinkDB\footnote{\small \url{http://blindb.org/}}, Verdict\footnote{\small \url{http://verdictdb.org/}}, Simba\footnote{\small \url{https://initialdlab.github.io/Simba/index.html}}, have been designed. However, as pointed out in~\cite{ChaudhuriDK17}, `` it seems impossible to have an approximate query processing system that supports the richness of SQL with significant saving of work while providing an accuracy guarantee that is acceptable to a broad set of application workloads.''

\item[(3)] {\em Approximation computing} is a recent computation technique that returns a possibly inaccurate result rather than a guaranteed accurate result from a system point of view.  It involves with multiple layers of the system stack from software to hardware to systems (such as approximate circuits, approximate storage and loop perforation), and can be used for applications where an approximate result is sufficient for its purpose~\cite{AgrawalCGGNOPSS16,Mittal16b}. Recently, a workshop on approximate computing across the stack has been usefully held for research on hardware, programming languages and compiler support  for approximate computing  since 2014.  (see \eg 2016\footnote{\small \url{http://approximate.computer/wax2016/}}, 2017\footnote{\small  \url{http://approximate.computer/wax2017/}} and 2018\footnote{\small \url{http://approximate.computer/wax2018/}}). Besides the various task oriented quality metrics,  the quality-energy trade-off is also  concerned for approximate computing. For instance, allowing only 5\% loss of classification accuracy can provide 50 times energy saving for clustering algorithm k-means~\cite{Mittal16b}.
\ei

In this article, we present the idea of approximate computation for efficient and effective big data analytics: query approximation and data approximation, based on our recent research experiences~\cite{MaCHW12,ShuaiMaVLDB12,tods-MaCFHW14,LinMZWH17,MaHWLH17,MaFLWCH16,MaFLWCH17,HuAMH16,DuanAMHH16,DuanMAMH17,rankicde2018}.
Approximation algorithms ask for feasible solutions that are theoretically bounded with respect to optimal solutions from an algorithm design aspect.
Approximate query processing and approximation computing relax the need for accuracy guarantees for aggregate SQL queries and for multiple layers of the system stack, respectively. Similarly, our approximate computation is unnecessarily theoretically bounded with respect to optimal solutions, but from an algorithm design point of view. That is, we focus on approximate computation for big data analytics for a situation where an approximate result is sufficient for a purpose.


\section{Query Approximation}
\label{sec-query}

\begin{figure}[tb!]
  \begin{center}
  \includegraphics[scale=0.45]{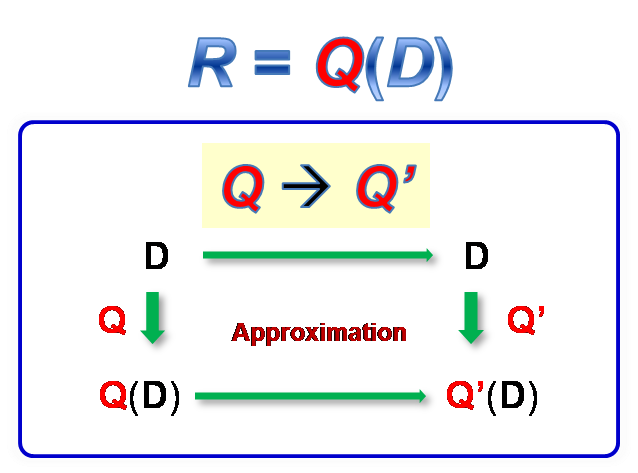}
  \caption{Query approximation}\label{fig-tech-queryappro}
  \end{center}
  \vspace{-1ex}
\end{figure}

Query approximation deals with complex queries involved with big data analytic tasks. Given a class $Q$ of data analytic queries with high a computational complexity,  query approximation is to transform into another class $Q'$ of queries with a low computational complexity and satisfiable approximate answers, as depicted in Fig.~\ref{fig-tech-queryappro} in which $Q$, $Q'$,  $D$ and $R$ denote the original query, approximate query, data and query result, respectively. Query approximation needs to reach a balance between the query efficiency and answer quality when approximating $Q$ with $Q'$.

The rationale behind query approximation lies in that inexact or approximate answers are sufficient or acceptable for many big data analytic tasks.
On one hand, when the volume of data is extremely large, it may be impossible or not necessary to compute the exact answers.
Observe that nobody would try each and every store to find a pair of shoes with the best cost-performance ratio.
That is, inexact (approximate) solutions are good enough for certain cases.
On the other hand, when taking noises (very common for big data) into account, it may not always be a good idea to compute exact answers
for those data analytic tasks whose answers are rare or hard to identify, such as the detection of homegrown violent extremists (HVEs) who seek to commit acts of terrorism in the United States and abroad~\cite{HungJ16}, as finding exact solutions may have a high chance to miss/ignore possible candidates.

We next explain query approximation computation in more detail using three different data analytic tasks.

\stitle{(1) Strong Simulation~\cite{ShuaiMaVLDB12,tods-MaCFHW14}}. Given a pattern graph $Q$ and a data graph $G$,
{\em graph pattern matching} is to find all subgraphs of $G$ that match $Q$, and is being increasingly used in various applications, \eg biology and social networks.

Here {\em matching} is typically defined in terms of
{\em subgraph isomorphism} \cite{Galla06}:
a subgraph $G_s$ of $G$ {\em matches} $Q$ if
there exists a {\em bijective function} $f$
from the nodes of $Q$ to the nodes in $G_s$ such that (a)  for each
pattern node $u$ in $Q$, $u$ and $f(u)$
have the same label,
and (b) there exists an edge $(u, u')$ in $Q$ if and only
if there exists an edge $(f(u), f(u'))$ in $G_s$.

The goodness of subgraph isomorphism is that all matched subgraphs  are exactly the same as the pattern graph, \ie completely preserving the  topology structure between the pattern graph and data graph. However, subgraph isomorphism is \NP-complete, and may return exponentially many matched subgraphs.
Further, subgraph isomorphism is too restrictive to find sensible matches in certain scenarios, as observed in~\cite{FanLMTWW10}. Even worse, online data in many cases only represents a partial world (\eg terrorist collaboration networks and homosexual networks are often accompanied with a large amount of offline data).
Exact computations on online data, whose offline counterpart is extremely hard to collect, typically decreases the chance of identifying candidate answers.
These hinder the usability of graph pattern matching in emerging applications.

To lower the high complexity of subgraph isomorphism, substitutes for subgraph isomorphism \cite{FanLMTWW10,FanLMTW11}, which allow graph pattern matching to be conducted in cubic-time, have been proposed by extending graph simulation~\cite{infsimu95}. However, they fall short of capturing the topology of data graphs, i.e., graphs may have a structure drastically different from pattern graphs that they match, and the matches found are often too large to analyze.

To rectify these problems, strong simulation, an ``approximate'' substitute for subgraph isomorphism, is proposed for graph pattern matching~\cite{tods-MaCFHW14}, which (a) theoretically preserves the key topology of pattern graphs and finds a bounded number of matches, (b) retains the same complexity as earlier extensions of graph simulation~\cite{FanLMTWW10,FanLMTW11}, by providing a cubic-time algorithm for strong simulation compuation, and (c) has the locality property that allows us to develop an effective distributed algorithm to conduct graph pattern matching on distributed graphs.

Strong simulation is experimentally verified that it is able to identify sensible matches that are not found by subgraph isomorphism, and it finds high
quality matches that retain graph topology. Indeed, 70\%-80\% of matches found by subgraph
isomorphism are retrieved by strong simulation. Further, strong simulation is over $100$ times faster than subgraph isomorphism, and has a bounded number of matches.

\stitle{(2) One-Pass Trajectory Compression~\cite{LinMZWH17}}.  Trajectory compression (\aka trajectory simplification) is to compress data points in a trajectory to a set of continuous line segments, and is commonly used  in practice.

The compression ratios of lossless methods are poor, and querying on the compressed data is time consuming due to the reconstruction of the original data \cite{Nibali:Trajic}. Hence, lossy techniques, which  provide approximate solutions with good compression ratios and bounded errors, are the mainstream.

Piece-wise line simplification (\lsa) comes from the computational geometry, whose target is to approximate a given finer piece-wise linear curve by another coarser piece-wise linear curve ({normally} a subset of the former), such that the maximum distance of the former from the later is bounded by a user specified constant (\ie error bound). It is widely used due to its distinct advantages: (a) simple and easy to implement, (b) no need extra knowledge and suitable for freely  moving  objects, and (c) bounded errors with good compression ratios.

\lsa algorithms fall into two categories: {\em optimal} and {\em approximate}.
Optimal methods\cite{Imai:Optimal} are to find the minimum number of points or segments to represent the original polygonal lines \wrt an error bound $\epsilon$. They have higher time and space complexities, and are not practical for large trajectory data.
Hence,  various approximate \lsa algorithms have been developed, from batch algorithms (\eg \cite{Douglas:Peucker}) to online algorithms (\eg~\cite{Liu:BQS}) and to one-pass algorithms (\eg~\cite{LinMZWH17}).

An \lsa algorithm is {\em one-pass} if it processes each point in a trajectory once and only once when compressing the trajectory.
Obviously, one-pass algorithms have low time and space complexities, and are more appropriate for online processing. The difficulty comes from the need to achieve effective compression ratios.
Existing trajectory simplification algorithms (\eg \cite{Douglas:Peucker}) and online algorithms  (\eg \cite{Liu:BQS}) essentially employ global distance checking, although online algorithms restrict the checking within a window. That is, whenever a new line segment is formed, these algorithms always check its distances to all or a subset of data points, and, therefore, a data point is checked multiple times, depending on its order in the trajectory and the number of directed line segments formed. Hence, {\em an appropriate local distance checking approach is needed in the first place for  one-pass trajectory simplification.}.

 We develop a local distance checking method, referred to as  {\em fitting function}, such that a data point is checked only once in the entire process. Based on the fitting function, we develop one-pass error bounded trajectory simplification algorithms \operb and \operba that scan each data point in a trajectory once and only once, allowing interpolating new data points or not, respectively. By comparing our algorithms with \fbqsa (the fastest existing \lsa online algorithm \cite{Liu:BQS}) and \dpa (the best existing \lsa batch algorithm in terms of compression ratio \cite{Douglas:Peucker}), our one-pass algorithms  \operb and \operba are over four times faster than \fbqsa, and have  comparable compression ratios with \dpa.

\stitle{(3) Dense Temporal Subgraph Computation~\cite{MaHWLH17}}.  We study dense subgraphs in {\em a special type of temporal networks} whose nodes and edges are kept fixed, but edge weights constantly and regularly vary with timestamps~\cite{MaHWLH17}.  Essentially, a temporal network with $T$ timestamps can be viewed as $T$ snapshots of a static network such that the network nodes and edges are kept the same among these $T$ snapshots, while the edge weights vary with network snapthots. Road traffic networks typically fall into this category of temporal networks, and dense subgraphs are used for road traffic analyses that are of particular importance for transportation management of large cities.

Dense subgraphs are a general concept, and their concrete semantics highly depends on the studied problems and applications. Though  dense subgraphs have been widely studied in static networks, how to properly define their semantics over temporal networks is still in the early stage, not to mention effective and efficient analytic algorithms.

We adopt the {\em form of dense temporal subgraphs} initially defined and studied in \cite{BogdanovMS11}, such that a temporal subgraph corresponds to a connected subgraph measured by the sum of all its edge weights in a time interval, \ie  a continuous sequence of timestamps. Intuitively, a dense subgraph that we consider  corresponds to a  collection of connected highly slow or jam roads (\ie  a jam area) in road networks, lasting for a continuous sequence of snapshots.

The problem of  finding dense subgraphs in temporal networks is non-trivial, and it is already \NP-complete even for a temporal network with a single snapshot and with $+1$ or $-1$ edge weights only, as observed in \cite{BogdanovMS11}. Even worse, it remains hard to approximate for temporal networks  with single snapshots~\cite{MaHWLH17}. Moreover, given a temporal network with $T$ timestamps, there are a total number of $T*(T+1)/2$ time intervals to consider, which further aggravates the difficulty. The state of the art solution \kw{MEDEN} \cite{BogdanovMS11} adopts a Filter-And-Verification ({\kw{FAV}) framework that {\em even if a large portion of time intervals are filtered, there often remain a large number of time intervals to verify}. Hence, this method is not big data friendly, and is not scalable when temporal networks have a large number of nodes/edges or a large number $T$ of timestamps.

We develop a data-driven approach (referred to as \kw{FIDES}), instead of filter-and-verification, to  identifying the most possible $k$ time intervals from $T \times (T + 1)/2$ time intervals, in which $T$ is the number of snapshots and k is a small constant, \eg 10. This is achieved by exploring the characteristics of time intervals involved with dense subgraphs based on the observation of {\em evolving convergence phenomenon} in traffic data, inspired by the convergent evolution in nature\footnote{\small \url{https://en.wikipedia.org/wiki/Convergent_evolution}}. That is, our method provides time intervals with probabilistic guarantees, instead of exact ones as \kw{FAV}.   Using both real-life and synthetic data, we experimentally show that our method \kw{FIDES} is over 1000 times faster than \kw{MEDEN}~\cite{BogdanovMS11}, while the quality of dense subgraphs found is comparable with  \kw{MEDEN} .

\section{Data Approximation}
\label{sec-data}

Big data has a large volume, and, hence, the space complexity~\cite{CormenLRS01} of big data analytic tasks starts raising more concerns.
Given a class $Q$ of queries on data $D$, data approximation is to transform $D$ into smaller $D'$ such that $Q$ on $D'$ returns a sufficient or satisfiable approximate answer in a more efficient way. Further, it is typically common that query $Q$ needs to be (slightly) modified to $Q'$ to accommodate  the changes of $D$ to $D'$, as shown in Fig.~\ref{fig-tech-dataappro}. Similar to query approximation, data approximation needs to reach a balance between the query efficiency and answer quality.

The rationale behind data approximation has roots in the Pareto principle\footnote{\small \url{https://en.wikipedia.org/wiki/Pareto_principle}} that ``states that, for many events, roughly 80\% of the effects come from 20\% of the causes''. The critical thing for data approximation is to determine which part of data is relevant to tasks (belong to the 20\%).
 By this principle, for many big analytic tasks, one may only need to keep a small amount of the data to derive high quality answers.
For example, when we are to build a predictive model on the stock of razers for an online store based on the order history of customers, orders from men are good enough. While on the stock of lipsticks, those from women are good enough. That is to say,  it is not necessary to use the entire data for certain data analytic tasks.

However, it should be pointed out that there are data analytic tasks such that data approximation could not work well. For example, an online store needs to count the total number of goods in its catalog.  Essentially entire goods should be considered for this task, and if a (small) portion of goods are chosen, it is hard to have a satisfiable result.

We next explain data approximation computation in more detail using three different data analytic tasks.

\begin{figure}[tb!]
  \begin{center}
  \includegraphics[scale=0.45]{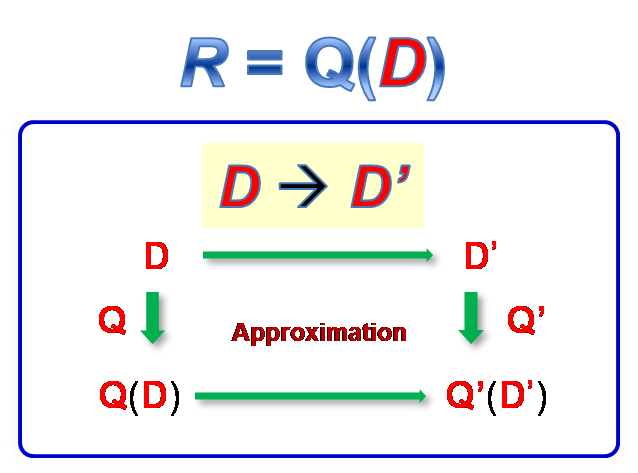}
  \end{center}
  \caption{Data approximation}\label{fig-tech-dataappro}
  \vspace{-1ex}
\end{figure}

\stitle{(1) Proxies for Shortest Paths and Distances~\cite{MaFLWCH16,MaFLWCH17}}. Computing shortest paths and distances is one of the fundamental problems on graphs. We study the {\em node-to-node shortest path} ({\em distance}) problem on large graphs: given a weighted undirected graph $G(V, E)$ with non-negative edge weights, and two nodes of $G$, the source $s$ and the target $t$, find the shortest path (distance) from $s$ to $t$ in $G$. The Dijkstra's algorithm with Fibonacci heaps runs in $O(n\log n + m)$ due to Fredman \& Tarjan~\cite{CormenLRS01}, where $n$ and $m$ denote the numbers of nodes and edges in a graph, respectively, which remains asymptotically the fastest known solution on arbitrary undirected graphs with non-negative edge weights.
However, computing shortest  paths and distances remains a challenging problem, in terms of both time and space cost, on large-scale graphs. Hence, various optimizations have been developed to speed-up the computation.

To speed-up shortest  path and distance queries, we propose {\em proxies} that have the following properties:
(a) each proxy captures a set of nodes in a graph, referred to as \dra,
(b) a small number of proxies can represent a large number of nodes in a graph,
(c) shortest paths and distances involved within the set of nodes being represented by the same proxies can be answered efficiently, and,
(d) the proxies and the set of nodes being represented can be computed efficiently in {\em linear} time.

The framework for speeding-up shortest path and distance queries with proxies consists of two module, preprocessing and query answering, as follows.

\ni(a) {\em Preprocessing}: Given graph $G(V, E)$, it first computes all \dras and their maximal proxies in linear time, then it computes and stores all the shortest paths and distances between any node and its proxy. It finally computes the reduced subgraph $G'$ by removing all \dras from graph $G$, \ie keeping the proxies only.

\ni(b) {\em Query answering}. Given two nodes $s$ and $t$ in graph $G(V$, $E)$  and the pre-computed information, the query answering module essentially executes the following.

The shortest path $\path(s, t)$ = $\path(s, u_s)/$ $\path(u_s, u_t)/$ $\path(u_t, t)$, where $u_s$ and $u_t$ are the proxies of $s$ and $t$, respectively.
As  $\path(s, u_s)$ and $\path(u_t, t)$ are pre-computed, and $\path(u_s, u_t)$ can be computed on the reduced subgraph $G'$ by invoking any existing algorithms
(\eg \ah~\cite{zhu2013shortest}).
The shortest distance $\dist(s, t)$ = $\dist(s, u_s)$ + $\dist(u_s, u_t)$ + $\dist(u_t, t)$ can be computed along the same line.

Essentially, we propose a light-weight data reduction technique for speeding-up (exact)  shortest path and distance queries on large weighted undirected graphs~\cite{MaFLWCH16}. We experimentally show that about $1/3$  nodes of real-life social and road networks  are captured by proxies.

\stitle{(2) Network Anomaly Detection~\cite{HuAMH16}}
Anomaly (or outlier) detection aims at identifying those objects in a dataset that are unusual, i.e., different than the majority and therefore suspicious resulting from a contamination, error, or fraud~\cite{Zimek2017}. Network anomaly detection has become very popular
recently because of the importance of discovering key
regions of structural inconsistency in the network. In addition
to application-specific information carried by anomalies, the
presence of such structural inconsistency is often an impediment
to the effective application of data mining algorithms such as
community detection and classification.

Networks are inherently complex entities, and, hence, anomalies may be defined in a wide variety of ways.
Our goal is to discover {\em  structural inconsistencies, i.e.,  the anomalous nodes that connect to a number of diverse influential communities}, inspired by the concept of social brokers across groups, which provide social capital in networks~\cite{s-hole04}.
While a variety of graph embeddings, such as multidimensional scaling~\cite{mds}, are available in the literature, they aim to preserve (global) pairwise similarities and are not optimized to networks and the problem of anomaly detection.
Hence, they cannot be directly used for the detection of structural inconsistencies proposed in this paper. We propose a novel graph embedding method, specifically designed to ferret out the anomalous nodes in large networks.

Our embedding approach is  based on a model, in which each dimension of the embedding corresponds to a clustered region in the network.
In other words, the similarity of different nodes along a particular dimension, indicates their similarity to a particular clustered region.
Therefore, this embedding retains a very high level of interpretability in  terms of the original graph data, which is very useful from an application-specific perspective. The nature of  the embedding also  makes it  possible  to detect anomalous nodes, by examining the interaction of each node with the different regions in terms of the embedding. In particular, we measure the {\em level of anomalousness} of a node in terms of the embedding imposed on the node and its neighbors.

Each node in graph embedding is represented as a d-dimensional vector, and the dimensionality $d$ can be large. Hence, such an approach is rather hard to apply to the case of large networks, because the complexity of the approach is in proportion to the square of the number of nodes when optimizing the embedding and because the noises in the embedding seriously impair the accuracy of detected anomalous nodes. Hence, we incorporate data approximation techniques (sampling, graph partitioning, and, moreover, a novel dimension reduction technique) to make the approach more scalable and effective for large networks.

Essentially, in our graph embedding, $d$ represents the number of communities. As the anomalous nodes are only determined by influential communities and nodes typically connect to a limited number of communities, the complete d-dimensions are unnecessary, and a limited number of communities suffice to ascertain anomalies. Thus, our dimension reduction technique (referred to as $k+\beta$ reduction) only maintains ($k + \beta$)-dimensions for embedding of each node, where k is the maximum number of communities to connect, $\beta$ is to tolerate mistakes when determining the $k$ communities and is removed after the computation process. Here $k, \beta \ll d$, \eg $k = 10$ and $\beta = 2$ and $d=600$ for a network with $10^6$ nodes.

Using both real-life data and synthetic
data , we conduct an extensive experimental study.
 (a) The modularity \cite{newman1} was increased about 4.9\% and 3.6\% with our approach and OddBall~\cite{akoglu14}, respectively; (c) Our approach scales to graph graphs with large number of communities, while traditional  multidimensional scaling approach~\cite{mds} ran out of memory.

\stitle{(3) Ensemble Enabled Link Prediction~\cite{DuanAMHH16,DuanMAMH17}}. Link prediction is the task to predict the formation of future links in a dynamic and evolving network, and has been extensively studied due to its numerous applications, such as the recommendation of friends in a social network, images in a multimedia network, or collaborators in a scientific network.

Link prediction methods are often applied to very large and sparse networks, which have a large {\em search space} $O(n^2)$,
where $n$ is the number of nodes. Hence, the scalability is a big challenge. In fact, an often overlooked fact is that most {\em exiting link prediction algorithms evaluate the link propensities only over a subset of possibilities rather than all propensities over the entire network}. 
Consider a large network with $10^8$ nodes. Its number of {\em possibilities} for links
is of the order of $10^{16}$. Therefore, a 3GHz processor would
require at least $35$ days just to allocate one {\em machine cycle} to
every pair of nodes. This implies that in order to determine the
top-ranked link predictions over the {\em entire network}, the
running time would be much more than $35$ days.

It is noteworthy that most existing link prediction algorithms are not designed to search over the entire
$O(n^2)$ possibilities. A closer examination of the relevant
studies shows that even for networks of modest size, these
algorithms perform benchmark evaluations over a {\em
sample of the possibilities} for links.  In other
words, the {\em complete ranking problem for link prediction in
very large networks remains challenging at least from a
computational point of view}.

Latent factor models have proven a great success for
collaborative filtering, but not link prediction in spite of the obvious
similarity and
the obvious effectiveness of latent factor models. One of the reasons why latent factor models are rarely used for
link prediction is due to their complexity. In
collaborative filtering applications, items have a few hundred thousand dimensions, whereas even the smallest  real-world networks contain more than a million nodes.
Even worse, we also experientially verify that the quality of link prediction for latent factor models decreases with the increase of data sparsity,
and networks typically become sparser when their sizes grow larger.

\eat{

The factorization of a matrix of
size $O(n^2)$ is not only computationally expensive, but also
memory-intensive.  As will be seen later in this article, one advantage
of  latent-factor models is that they are able to  transform the
adjacency matrix to a multidimensional space which can be searched
efficiently {\em by pruning} large portions of the $O(n^2)$ search
space in order to recommend the top-$k$ possibilities.

\begin{table}
\caption{The $O(n^2)$ problem in link prediction: Time required to
allocate a {\em single machine cycle} to every node-pair possibility
in networks of varying sizes and processors of various speeds.}
\label{time}
\vspace{0ex}
\centering
\begin{tabular}{cccc}
\hline \hline Network Sizes & 1 GHz &  3 GHz & 10 GHz \\
\hline \hline $10^6$ nodes & 1000 sec. & 333 sec. & 100 sec.\\
\hline $10^7$ nodes & 27.8 hrs &  9.3 hrs &  2.78 hrs\\
\hline $10^8$ nodes & $>100$ days &  $>35$ days & $> 10$ days\\
\hline $10^9$ nodes & $>10000$ days & $>3500$ days & $> 1000$ days\\
\hline \hline
\end{tabular}
\vspace{-2ex}
\end{table}
}

We explore an {\em ensemble approach} to making latent factor models
practical for link prediction by decomposing the search space into a
 set of smaller matrices with three structural bagging methods with performance guarantees, which has obvious {\em
effectiveness} advantages. In this way, latent factor models only need to deal with networks with small sizes (and denser), and retain
their effectiveness and efficiency.  By incorporating with the characteristics of  link prediction, the bagging methods maintain high prediction
accuracy while reducing the network size via graph sampling techniques.
Further, the use of an ensemble
approach has obvious robustness advantages as well.

We experimentally show that our ensemble approach is over $50$ times faster and  over $20\%$ more accurate than {\sc bigclam} \cite{yang-wsdm2013} using real-life social networks.

\eat{
\stitle{(1) Network Anomaly Detection~\cite{HuAMH16}}
We have adopted the idea in the process of dealing with large graphs in the study of anomaly detection in graph streams, when dealing with the matrix representation of a social graph, and  we have both theoretically and experimentally shown that simplifying the matrix by replacing a part of small entry values  with zero has few affects on the computation of eigenvectors~\cite{YuAMW13}.
}

\section{Beyond Approximation Techniques}
\label{sec-beyond}

For big data analytics, there are no one-size-fits-all techniques, and it is often necessary to combine different techniques to obtain good solutions.

We have seen that sampling helps to achieve a balance between efficiency and effectiveness for approximate query processing~\cite{ChaudhuriDK17,Mozafari17,Kraska17,GarofalakisG01} and link prediction~\cite{DuanAMHH16,DuanMAMH17}},
and other techniques such as incremental computation~\cite{Reps96,FanLMTWW10,rankicde2018}, distributed computing~\cite{MaCHW12,FanXWYJZZCT17}, and system techniques~\eg caching~\cite{WangLMNT18}, hardware~\cite{AbergerLTNOR17,Han0Y18} can also be unitized for big data analytics, and can even be combined for designing query and data approximation techniques for big data analytics. 

It is worth pointing out that (1) for all kind of techniques big data analytics, various computing resources should be seriously considered, \eg using bounded resources for approximation~\cite{CaoF17} and for incremental computation~\cite{abs-1801-01012}, and (2) theoretical analyses are also important for developing approximation techniques. For instance, our query and data approximation techniques are based serious theoretical results~\cite{ShuaiMaVLDB12,tods-MaCFHW14,LinMZWH17,MaFLWCH16,MaFLWCH17}.

\section{Conclusions}
\label{sec-conclusion}

In this article we have systematiclly introduced approximation computation techniques for efficient and effective big data analytics.
Furthermore, although approximate computation does not put
theoretical bounds with respect to optimal solutions, it does expect a balance between efficiency and effectiveness. Indeed, (a) our query approximation techniques~\cite{ShuaiMaVLDB12,tods-MaCFHW14,LinMZWH17,MaHWLH17} show that efficiency can be obtained with high accuracy in practice, and (b) our data approximation techniques~\cite{MaFLWCH16,MaFLWCH17,HuAMH16,DuanAMHH16,DuanMAMH17} show that efficiency and accuracy can be obtained simultaneously for certain data analytic tasks. That is, though approximate computation is for a situation where an approximate result is sufficient for a purpose, its design policy is not always to sacrifice effectiveness for efficiency.



\stitle{Acknowledgement}. This work is supported in part by 973 program ({\small 2014CB340300}), NSFC ({\small U1636210 \& 61421003}). We would also thank our colleagues Charu Aggarwal, Wenfei Fan, Xuelian Lin and our students Yang Cao, Liang Duan, Kaiyu Feng, Renjun Hu, Han Zhang for their joined efforts.

\newpage
\balance
\bibliographystyle{abbrv}
\vspace{-1ex}
\begin{small}
\bibliography{paper}

\begin{thebibliography}{10}

\bibitem{AbergerLTNOR17}
C.~R. Aberger, A.~Lamb, S.~Tu, A.~N{\"{o}}tzli, K.~Olukotun, and C.~R{\'{e}}.
\newblock Emptyheaded: {A} relational engine for graph processing.
\newblock {\em {ACM} Trans. Database Syst.}, 42(4):20:1--20:44, 2017.

\bibitem{AgrawalCGGNOPSS16}
A.~Agrawal, J.~Choi, K.~Gopalakrishnan, S.~Gupta, R.~Nair, J.~Oh, D.~A. Prener,
  S.~Shukla, V.~Srinivasan, and Z.~Sura.
\newblock Approximate computing: Challenges and opportunities.
\newblock In {\em {ICRC}}, 2016.

\bibitem{akoglu14}
L.~Akoglu, H.~Tong, and D.~Koutra.
\newblock Graph based anomaly detection and description: a survey.
\newblock {\em Data Min. Knowl. Discov.}, 29(3):626--688, 2015.

\bibitem{Ausiello99}
G.~Ausiello, P.~Crescenzi, G.~Gambosi, V.~Kann, Marchetti-Spaccamela, and
  M.~A., Protasi.
\newblock {\em Complexity and Approximation: Combinatorial Optimization
  Problems and Their Approximability Properties}.
\newblock Springer, 1999.

\bibitem{BogdanovMS11}
P.~Bogdanov, M.~Mongiov{\`{\i}}, and A.~K. Singh.
\newblock Mining heavy subgraphs in time-evolving networks.
\newblock In {\em {ICDM}}, 2011.

\bibitem{mds}
I.~Borg and P.~Groenen.
\newblock {\em Modern Multidimensional Scaling: Theory and Applications (2nd
  ed.)}.
\newblock Springer, 2005.

\bibitem{s-hole04}
R.~S. Burt.
\newblock Structural holes and good ideas.
\newblock {\em American Journal of Sociology}, 110(2):349--399, 2004.

\bibitem{CaoF17}
Y.~Cao and W.~Fan.
\newblock Data driven approximation with bounded resources.
\newblock {\em {PVLDB}}, 10(9):973--984, 2017.

\bibitem{ChaudhuriDK17}
S.~Chaudhuri, B.~Ding, and S.~Kandula.
\newblock Approximate query processing: No silver bullet.
\newblock In {\em {SIGMOD}}, 2017.

\bibitem{newman1}
A.~Clauset, M.~E.~J. Newman, and C.~Moore.
\newblock Finding community structure in very large networks.
\newblock {\em Physical Review E}, 70:066111, 2004.

\bibitem{CormenLRS01}
T.~H. Cormen, C.~E. Leiserson, R.~L. Rivest, and C.~Stein.
\newblock {\em Introduction to Algorithms}.
\newblock The MIT Press, 2001.

\bibitem{Douglas:Peucker}
D.~H. Douglas and T.~K. Peucker.
\newblock Algorithms for the reduction of the number of points required to
  represent a digitized line or its caricature.
\newblock {\em The Canadian Cartographer}, 10(2):112--122, 1973.

\bibitem{DuanAMHH16}
L.~Duan, C.~C. Aggarwal, S.~Ma, R.~Hu, and J.~Huai.
\newblock Scaling up link prediction with ensembles.
\newblock In {\em {WSDM}}, 2016.

\bibitem{DuanMAMH17}
L.~Duan, S.~Ma, C.~Aggarwal, T.~Ma, and J.~Huai.
\newblock An ensemble approach to link prediction.
\newblock {\em {TKDE}}, 29(11):2402--2416, 2017.

\bibitem{FanGN13}
W.~Fan, F.~Geerts, and F.~Neven.
\newblock Making queries tractable on big data with preprocessing.
\newblock {\em {PVLDB}}, 6(9):685--696, 2013.

\bibitem{FanLMTW11}
W.~Fan, J.~Li, S.~Ma, N.~Tang, and Y.~Wu.
\newblock Adding regular expressions to graph reachability and pattern queries.
\newblock In {\em ICDE}, 2011.

\bibitem{FanLMTWW10}
W.~Fan, J.~Li, S.~Ma, N.~Tang, Y.~Wu, and Y.~Wu.
\newblock Graph pattern matching: From intractable to polynomial time.
\newblock {\em PVLDB}, 3(1), 2010.

\bibitem{FanXWYJZZCT17}
W.~Fan, J.~Xu, Y.~Wu, W.~Yu, J.~Jiang, Z.~Zheng, B.~Zhang, Y.~Cao, and C.~Tian.
\newblock Parallelizing sequential graph computations.
\newblock In {\em {SIGMOD}}, 2017.

\bibitem{Galla06}
B.~Gallagher.
\newblock Matching structure and semantics: A survey on graph-based pattern
  matching.
\newblock {\em AAAI FS.}, 2006.

\bibitem{GareyGU72}
M.~R. Garey, R.~L. Graham, and J.~D. Ullman.
\newblock Worst-case analysis of memory allocation algorithms.
\newblock In {\em {STOC}}, 1972.

\bibitem{GarofalakisG01}
M.~N. Garofalakis and P.~B. Gibbons.
\newblock Approximate query processing: Taming the terabytes.
\newblock In {\em {VLDB}}, 2001.

\bibitem{Han0Y18}
S.~Han, L.~Zou, and J.~X. Yu.
\newblock Speeding up set intersections in graph algorithms using {SIMD}
  instructions.
\newblock In {\em {SIGMOD}}, 2018.

\bibitem{infsimu95}
M.~R. Henzinger, T.~A. Henzinger, and P.~W. Kopke.
\newblock Computing simulations on finite and infinite graphs.
\newblock In {\em FOCS}, 1995.

\bibitem{Dorit96}
D.~S. Hochbaum.
\newblock {\em Approximation Algorithms for NP-Hard Problems}.
\newblock Springer, 1996.

\bibitem{HuAMH16}
R.~Hu, C.~C. Aggarwal, S.~Ma, and J.~Huai.
\newblock An embedding approach to anomaly detection.
\newblock In {\em {ICDE}}, 2016.

\bibitem{HungJ16}
B.~W.~K. Hung and A.~P. Jayasumana.
\newblock Investigative simulation: Towards utilizing graph pattern matching
  for investigative search.
\newblock In {\em {ASONAM}}, 2016.

\bibitem{Imai:Optimal}
H.~Imai and M.~Iri.
\newblock Computational-geometric methods for polygonal approximations of a
  curve.
\newblock {\em Computer Vision, Graphics, and Image Processing}, 36:31--41,
  1986.

\bibitem{Johnson74a}
D.~S. Johnson.
\newblock Approximation algorithms for combinatorial problems.
\newblock {\em J. Comput. Syst. Sci.}, 9(3):256--278, 1974.

\bibitem{Jordan15}
M.~I. Jordan.
\newblock Computational thinking, inferential thinking and ``big data''.
\newblock In {\em {PODS}}, 2015.

\bibitem{Kraska17}
T.~Kraska.
\newblock Approximate query processing for interactive data science.
\newblock In {\em {SIGMOD}}, 2017.

\bibitem{LinMZWH17}
X.~Lin, S.~Ma, H.~Zhang, T.~Wo, and J.~Huai.
\newblock One-pass error bounded trajectory simplification.
\newblock {\em {PVLDB}}, 10(7):841--852, 2017.

\bibitem{Liu:BQS}
J.~Liu, K.~Zhao, P.~Sommer, S.~Shang, B.~Kusy, and R.~Jurdak.
\newblock Bounded quadrant system: Error-bounded trajectory compression on the
  go.
\newblock In {\em ICDE}, 2015.

\bibitem{ShuaiMaVLDB12}
S.~Ma, Y.~Cao, W.~Fan, J.~Huai, and T.~Wo.
\newblock Capturing topology in graph pattern matching.
\newblock {\em {PVLDB}}, 5(4):310--321, 2011.

\bibitem{tods-MaCFHW14}
S.~Ma, Y.~Cao, W.~Fan, J.~Huai, and T.~Wo.
\newblock Strong simulation: Capturing topology in graph pattern matching.
\newblock {\em {TODS}}, 39(1):4:1--4:46, 2014.

\bibitem{MaCHW12}
S.~Ma, Y.~Cao, J.~Huai, and T.~Wo.
\newblock Distributed graph pattern matching.
\newblock In {\em {WWW}}, 2012.

\bibitem{MaFLWCH16}
S.~Ma, K.~Feng, J.~Li, H.~Wang, G.~Cong, and J.~Huai.
\newblock Proxies for shortest path and distance queries.
\newblock {\em {TKDE}}, 28(7):1835--1850, 2016.

\bibitem{MaFLWCH17}
S.~Ma, K.~Feng, J.~Li, H.~Wang, G.~Cong, and J.~Huai.
\newblock Proxies for shortest path and distance queries.
\newblock In {\em {ICDE}}, 2017.

\bibitem{rankicde2018}
S.~Ma, C.~Gong, R.~Hu, D.~Luo, C.~Hu, and J.~Huai.
\newblock Query independent scholarly article ranking.
\newblock In {\em {ICDE}}, 2018.

\bibitem{MaHWLH17}
S.~Ma, R.~Hu, L.~Wang, X.~Lin, and J.~Huai.
\newblock Fast computation of dense temporal subgraphs.
\newblock In {\em {ICDE}}, 2017.

\bibitem{abs-1801-01012}
S.~Ma, J.~Li, C.~Hu, X.~Liu, and J.~Huai.
\newblock Graph pattern matching for dynamic team formation.
\newblock {\em CoRR}, abs/1801.01012, 2018.

\bibitem{Mittal16b}
S.~Mittal.
\newblock A survey of techniques for approximate computing.
\newblock {\em {ACM} Comput. Surv.}, 48(4):62:1--62:33, 2016.

\bibitem{Mozafari17}
B.~Mozafari.
\newblock Approximate query engines: Commercial challenges and research
  opportunities.
\newblock In {\em {SIGMOD}}, 2017.

\bibitem{Nibali:Trajic}
A.~Nibali and Z.~He.
\newblock {Trajic}: An effective compression system for trajectory data.
\newblock {\em TKDE}, 27(11):3138--3151, 2015.

\bibitem{Reps96}
G.~Ramalingam and T.~Reps.
\newblock On the computational complexity of dynamic graph problems.
\newblock {\em TCS}, 158(1-2), 1996.

\bibitem{approx03}
V.~V. Vazirani.
\newblock {\em Approximation Algorithms}.
\newblock Springer, 2003.

\bibitem{WangLMNT18}
J.~Wang, Z.~Liu, S.~Ma, N.~Ntarmos, and P.~Triantafillou.
\newblock {GC:} {A} graph caching system for subgraph/supergraph queries.
\newblock {\em {PVLDB}}, 11(12):2022--2025, 2018.

\bibitem{yang-wsdm2013}
J.~Yang and J.~Leskovec.
\newblock Overlapping community detection at scale: A nonnegative matrix
  factorization approach.
\newblock In {\em WSDM}, 2013.

\bibitem{ZahariaXWDADMRV16}
M.~Zaharia, R.~S. Xin, P.~Wendell, T.~Das, M.~Armbrust, A.~Dave, X.~Meng,
  J.~Rosen, S.~Venkataraman, M.~J. Franklin, A.~Ghodsi, J.~Gonzalez,
  S.~Shenker, and I.~Stoica.
\newblock Apache spark: a unified engine for big data processing.
\newblock {\em Commun. {ACM}}, 59(11):56--65, 2016.

\bibitem{zhu2013shortest}
A.~D. Zhu, H.~Ma, X.~Xiao, S.~Luo, Y.~Tang, and S.~Zhou.
\newblock Shortest path and distance queries on road networks: towards bridging
  theory and practice.
\newblock In {\em SIGMOD}, 2013.

\bibitem{Zimek2017}
A.~Zimek and E.~Schubert.
\newblock {\em Outlier Detection}, pages 1--5.
\newblock Springer New York, 2017.

\end{thebibliography}
\end{small}

\vspace{-4ex}
\begin{IEEEbiography}
{Shuai Ma} is a professor at the School of Computer Science and Engineering, Beihang University, China.
He obtained his PhD degrees from University of Edinburgh in 2010, and from
Peking University in 2004, respectively.
He was a postdoctoral research fellow in the database group, University of Edinburgh, a summer intern at Bell labs, Murray Hill, USA and a visiting researcher of MRSA.
He is a recipient of the best paper award for VLDB 2010 and the best challenge paper award for WISE 2013. His current research interests include database theory and systems, social data and graph analysis, and data intensive computing.
\end{IEEEbiography}
\vspace{-4ex}
\begin{IEEEbiography}
{Jinpeng Huai} is a professor at the School of Computer Science and Engineering, Beihang University, China. He received his Ph.D. degree in computer science from Beihang University, China, in 1993. Prof. Huai is an academician of Chinese Academy of Sciences and the vice honorary chairman of China Computer Federation (CCF). His research interests include big data computing, distributed systems, virtual computing, service-oriented computing, trustworthiness and security.
\end{IEEEbiography}


\end{document}